\documentclass[11pt, a4paper]{article}

\usepackage{fancyhdr,amsmath,amsfonts,theorem}%

\usepackage{tikz} 
\usetikzlibrary{arrows}
\usepackage{caption}
\usepackage{longtable}
\usepackage{booktabs}
\usepackage{array}
\newcommand{\bigCI}{\mathrel{\text{\scalebox{1.07}{$\perp\mkern-10mu\perp$}}}}
\graphicspath{ {Images/} }
\tikzstyle{vertex}=[circle, fill,draw, inner sep=0pt, minimum size=6pt]
\newcommand{\vertex}{\node[vertex]}
\usepackage[colorlinks=true, citecolor=blue,linkcolor=blue]{hyperref}
\usepackage[ natbib=true, backend=bibtex]{biblatex}
\author{Niharika Gauraha \\ Indian Statistical Institute}
\title{ Graphical Log-linear Models: Fundamental Concepts and Applications}

\newtheorem{theorem1}{Special Theorem}
\newtheorem{definition}[theorem1]{Definition}
\newtheorem{proposition}[theorem1]{Proposition}

\newtheorem{example}[theorem1]{Example}

\begin{document}
\maketitle
Abstract:
We present a comprehensive study of graphical log-linear models for contingency tables. High dimensional contingency tables arise in many areas such as computational biology, collection of survey and census data and others. Analysis of contingency tables involving several factors or categorical variables is very hard. To determine interactions among various factors, graphical and decomposable log-linear models are preferred. First, we explore connections between the conditional independence in probability and graphs; thereafter we provide a few illustrations to describe how graphical log-linear model are useful to interpret the conditional independences between factors. We also discuss the problem of estimation and model selection in decomposable models. 

\section{Introduction}
In this paper, our aim is to provide the reader with insight into the graphical log-linear models by providing a concise explanation of the underlying mathematics and statistics, by pointing out relationships to conditional independence in probability and graphs and providing pointers to available software and important references.

Log Linear Models(abbreviated as LLMs) are the most widely used models for analyzing cross-classified categorical data, see \cite{Christensen}. Though LLM supports various range of models based on non-interaction assumptions. But for fairly large dimensional tables the analysis becomes difficult, as the number of factors increases the number of interaction terms grows exponentially. 
Graphical Log Linear Models(abbreviated as GLLMs) are a way of representing relationships among the factors of a contingency table using a graph. 
GLLMs have two great advantages: from the graph structure it is easy to read off the conditional independence relations and 
graph based algorithms usually provide efficient computational algorithms for parameter estimation and model selection.


The decomposable log linear models are a restricted class of GLLMs which are based on chordal graphs.  
There are several reasons for using decomposable models over an ordinary GLLM. Firstly, the maximum likelihood estimates can be found explicitly. Secondly, closed form expressions exists for test statistics. Another advantage is that it has triangulated graph based, efficient inference algorithms. Thus decomposable models are mostly used for analysis of high dimensional tables.

We have organized the rest of the article in the following manner. In section 2, we briefly review graph theory, conditional independence in probability and Markov Networks. Section 3 gives overview of contingency tables and describes different types of contingency tables based on the underlying sampling distributions. Section 4 introduces the theory of log-linear interaction models and defines various classes of LLMs such as comprehensive, hierarchical, graphical and decomposable LLMs. Section 5 is concerned with statistical properties of LLMs such as the sufficient statistics, the Maximum Likelihood Estimates(MLE) and model testing. In section 6, we discuss the analysis of three-factor contingency tables. In section 7, the backward model selection algorithm for decomposable models is illustrated with an example. Section 8 gives computational details. We shall provide some concluding remarks in section 9. 
\section{Graph Theory and Markov Networks}
In this section, we briefly review the graph theoretic concepts, the conditional independence in probability and Markov Networks.
\subsection{Graph Theory}
Here we list and define the necessary concepts of graph theory that we will be using in later sections. See \cite{West}, for further details on graph theory.

A graph G, is a pair G = (V, E), where V is a set of vertices and E is a set of edges. A graph is  said to be an undirected graph when E is a set of unordered pairs of vertices. We consider only simple graph that has neither loops nor multiple edges. 

\begin{definition}[Boundary]
Let $G = (V, E)$ be an undirected graph. The neighbours or boundary of a subset $A$ of vertices is a subset $C$ of vertices such that all nodes in C are not in A but are adjacent to some vertex in A.
\begin{align*}
	bd(A) = \{ u \in V \setminus A \mid \exists v \in A:\{u,v\} \in E \}
\end{align*}
\end{definition}

\begin{definition}[Maximal Clique]
A clique of a graph G is a subset C of vertices such that all vertices in C are mutually adjacent. A clique is said to be maximal if no vertex can be added to C without violating clique property.
\end{definition}
\begin{definition}[Chordal(Triangulated) Graphs]
In graph theory, a chord of a cycle C is defined as the edge which is not in the edge set of C but joins two vertices from the vertex set C. A graph is said to be chordal graph if every cycle of four or more length has a chord.
\end{definition}
\begin{definition}[Isomorphic Graphs]
Two graphs are said to be isomorphic if they have same number of vertices, same number of edges and they are connected in the same way.
\end{definition}

\subsection{Conditional Independence}
The concept of conditional independence in probability theory is very important and it is the basis for the graphical models. It is defined as follows.
 
\begin{definition} [Conditional Independence] Let $X, Y$ and $Z $ be random variables with a joint distribution P. The random variables X and Y are said to be conditionally independent given the random variable Z if and only if the following holds.
\begin{align*}
	 P(X,Y \mid Z) &= P(X \mid Z)P(Y \mid Z) \\
	 P(X \mid Y,Z)& = P(X \mid Z) 
\end{align*}
	 We sometimes also use David's notation (see \cite{Dawid} for details),  $ X \bigCI Y \mid Z $.
\end{definition}
Conditional independence has a vast literature in the field of probability and statistics, we refer to \cite{Dawid} and \cite{PearlPaz} for further details. 

\subsection{Markov Networks and Markov Properties}
In this section, we define Markov network graphs, Markov networks and different Markov properties for the Markov Networks. 

\begin{definition}[Markov Network Graphs]
A Markov network graph is an undirected graph G = ( V, E ), where $V= \{ X_1, X_2,..,X_n \}$ represents random variables of a multivariate distribution. 
\end{definition}

\begin{definition}[Markov Networks] \label{definition:6}
A Markov network M, is a pair $M = (G, \psi)$. Where G is a Markov network graph and $\psi = \{ \psi_1, \psi_2, ..., \psi_m\}$ is a set of non negative functions for each maximal clique	$C_i \in G $ $\forall i = 1 \dots m $ and the joint pdf can be decomposed into factors as
\begin{align*}
	P(x) =\frac{1}{Z} \prod_{a \in C_m} \psi_a(x)
\end{align*}
where Z is a normalizing constant.
\end{definition}

\begin{definition}[(P) Pairwise Markov Property] 
A probability distribution P satisfies the pairwise Markov property for a given undirected graph G if, for every pair of non adjacent edges X and Y , X is independent of Y given the rest. 
\begin{align*}
	 X \bigCI Y \mid (V \setminus {X,Y}) 
\end{align*}
\end{definition}
\begin{definition}[(L) Local Markov Property]
A probability distribution P satisfies the local Markov property for a given undirected graph G if, every variable X is conditionally independent of its non neighbours in the graph, given its neighbours. 
\begin{align*}
	X \bigCI (V \setminus {X \cup bd(X)}) \mid bd(X)
\end{align*}
where bd(X) denotes boundary of X.
\end{definition}
\begin{definition}[(G) Global Markov Property]
A probability distribution P, is said to be global Markov with respect to an undirected graph G if, for any disjoint subsets of nodes A, B, C such that C separates A and B on the graph, if and only if the distribution satisfies
\begin{align*}
	A \bigCI B | C
\end{align*}
\end{definition}

We must note that the above three Markov properties are not equivalent to each other. The Local Markov property is stronger than the pairwise one, while weaker than the global one. More precisely, we have the following proposition.
\begin{proposition}
For any probability measure the following holds.
\begin{align*}
	(G) \implies (L) \implies (P)
\end{align*}
\end{proposition}

See \cite{LaurtBook} for the proof of the proposition (11). We refer to \cite{Whittaker}, \cite{LaurtBook} and \cite{Edwards} for more details on graphical models and to \cite{LaurLLM1} and \cite{Darroch1} for Markov fields for LLMs.

\section*{Notations and Assumptions}
In this section, we discuss the notations and the assumptions which we will be using throughout the remaining sections of this article. We mainly consider the three-dimensional tables for notational simplification, which is also a true representative of k-dimensions and thus can be easily extended to any higher dimensions by increasing the number of subscripts. We mostly follow the notation from \cite{Christensen} and \cite{BishopBook}, for additional details we refer to these books.

Let us consider a three dimensional table with factors X, Y and Z. It must be noted that we interchangeably use numeric$\{1,2,3\}$ and alphabetic$\{X,Y,Z\}$ symbols for representing the factors of a contingency table. Suppose the factors X, Y and Z have I,J and K levels respectively, then we have an $I \times J \times K$ contingency table. \\
The following notations are defined for each elementary cell $(i,j,k) \; \; \; \forall i =  1 \ldots I, \forall j =  1 \ldots J, \forall k =  1 \ldots K $ 
\begin{align*}
	n_{ijk} &= \text{ the observed counts in the cell }(i,j,k)\\
	m_{ijk} &= \text{ the expected counts in the cell }(i,j,k)\\
	\hat{m_{ijk}} &= \text{ The Maximum Likelihood Estimate of }  m_{ijk}\\
	p_{ijk} &= \text{ probability of a count falling in cell }(i,j,k)\\
	\hat{p_{ijk}} &= \text{The Maximum Likelihood Estimate of } p_{i,j,k}
\end{align*}
The following notations are used for sums of elementary cell counts. where ``." represents summation over that factor. For example
\begin{align*}
	n_{i..} &= \sum_{jk} n_{ijk} = \sum_k n_{i.k}\\
	N = n_{...} &= \text{ total number of observations }
\end{align*}
 Similarly the marginal totals of probabilities and the expected counts are denoted by $p_{.jk}, $ and $m_{.jk}$ etc.

We represent ``C" as tables of sums obtained by summing over one or more factors, i.e. $C_{12}$ represents tables of counts $\{n_{ij.}\}$. Subscripted u-terms notation are used for main effects and interactions. For example  $u_{12(ij)}$ is used for two-factor interactions $  \forall i =  1 \ldots I, \forall j =  1 \ldots J$. We may interchangeably use $u_{12(ij)}$ and $u_{12}$, later one is obtained by simply dropping the second set of subscript. Thus $u_{12} = u_{12(ij)}  \forall i =  1 \ldots I, \forall j =  1 \ldots J$.

We assume that the observed cell counts are strictly positive for all models we consider throughout this article.
\section{Overview of Contingency Tables}
In this section, we  briefly review structural representation for count data called contingency tables. A contingency table is a table of counts that summarizes relationship between factors. In a multivariate qualitative data where each individual is described by a set of attributes, all individual with same attributes are counted and this count is entered into cell of a corresponding contingency table, see \cite{BishopBook}. The term ``contingency" was introduced by  \cite{Pearson}.  There is an extensive literature on contingency tables, see \cite{Andersen}, \cite{Bartlett} and \cite{Goodman1}.

\begin{example}[Example of a three-dimensional contingency table] \label{ex:1}
 example 3.2.1 of \cite{Christensen}.
\end{example}
	\begin{longtable}[h!] {@{}ccc|cc@{}} 
		\caption{Personality Type Table} \label{table:1}\\
		\toprule \centering
	 	&& &\multicolumn{2}{c}{Diastolic Blood Pressure}  \\
		\cmidrule{4-5}
		 Personality Type & \phantom{ab}& Cholestrol & Normal & High \\ \midrule
		
		A &&Normal& 716&79\\
		& &High&    207 & 25\\

		B &&Normal& 819 &67\\
		& &High&    186 & 22\\
		\bottomrule
	\end{longtable}

\subsection{Types of Contingency Tables}
Based on the underlying assumption of sampling distributions, contingency tables are divided into three main categories as follows.
\subsubsection{The Poisson Model}
In this model, it is assumed that cell counts are independent and Poisson-distributed. The total number of counts and the marginal counts are random variables. 
For three-dimensional tables with counts as random variables as $n_{ijk}$, the joint probability density function(pdf) can be written as \\
\begin{align} \label{eq:1}
	\prod_{ijk} \frac{m_{ijk}^{n_{ijk}} e^{-m_{ijk}}}{{ n_{ijk}!}}
\end{align}
\subsubsection{The Multinomial Model}
In this model, it is assumed that total number of subjects, N, are fixed. With this constraint imposed on independent Poisson distributions, the cell counts yield a multinomial distribution. For proof we refer to \cite{Fisher}. 

The pdf for this model is given as
\begin{align} \label{eq:2}
	\frac{N!}{ \prod_{ijk} n_{ijk}!} \prod_{ijk} \left(\frac{m_{ijk}}{N} \right)^{n_{ijk}}	
\end{align}

\subsubsection{The Product Multinomial Model}
In this model, it is assumed that one set of marginal count is fixed and The corresponding table of sums follow a product-multinomial distribution. For example, consider a three-dimensional table with total counts for factor 1, $n_{.jk}$, fixed. The pdf is given as \begin{align} \label{eq:3}
	\prod_{jk} \left[ \frac{n_{.jk}!}{ \prod_i n_{ijk}! } \prod_i \left( \frac{m_{ijk}}{n_{.jk}} \right)^{n_{ijk}} \right]
\end{align}

\section{Introduction to Log-linear Models}
This section introduces log-linear models for contingency tables. As discussed previously, the distribution of cell probabilities belong to exponential family(Poisson, multinomial and product-multinomial). Here we construct a linear model in the log scale of the expected cell count.\\
\\
A log-linear model for three-factor table is define as 
\begin{align} \label{eq:4}
	 \ln m_{ijk} &= u + u_{1(i)} + u_{2(j)} +u_{3(k)}+ u_{12(ij)} + u_{13(ik)} + u_{23(jk)} + u_{123(ijk)}
\end{align}
 with the following identifiability constraints:
\begin{align*}	 	 
	 \sum_i u_{1(i)} &= \sum_j u_{2(j)} = \sum_k u_{3(k)} = 0 \\
	 \sum_i u_{12(ij)} &= \sum_j u_{12(ij)} = 0 \\
	 \sum_i u_{13(ik)} &= \sum_k u_{13(ik)} = 0  \\
	 \sum_j u_{23(jk)} &= \sum_k u_{23(jk)} = 0 \\
	 \sum_i u_{123(ijk)} &= \sum_j u_{123(ijk)} = \sum_k u_{123(ijk)} = 0 
\end{align*}

The above model is called \textit{saturated} or \textit{unrestricted} because it contains all possible one-way, two-way and three-way effects. In general if no interaction terms are set to zero, it is called the saturated model. 

We must note that the number of terms in a log-linear model depends on the dimensions or number of factors and the interdependencies between the factors, it does not depend on the number of cells, see  \cite{Birch} for more details.

The model given by the equation(\ref{eq:4}) applies to the all three kinds of contingency tables with three factors(as discussed in previous section) but there may be differences in the interpretations of the interaction terms, see  \cite{Kreiner2} and \cite{Lang4}. There is a wide literature on LLMs, see for instance, \cite{Agresti}, \cite{Christensen}, \cite{Daniel}, \cite{BishopBook} and \cite{Knoke}.

\subsection{Log-Linear Models as Generalized Linear Models}
Let us recall the generalized linear model(GLM). It consists of a linear predictor and a link function. Link function determines relationship between the mean and the linear predictor. Here we show that the LLMs are special instances of GLMs for Poisson-distributed data, see \cite{Nelder} for the details.

Consider a $2 \times 2$ Poisson model with two factors say X and Y, suppose cell counts  $n_{ij}$ are response variables such that $ n_{ij} \sim Poisson(m_{ij})$ and the factors X and Y are explanatory variables. \\
define a link function  g as
\begin{align*}	
	g(m_{ij}) &= \ln m_{ij}
\end{align*}
the linear predictor is defined as $X^{'} \beta $ \\
where $X$ is the design matrix and $\beta$ is the vector of unknown parameters.\\
For this model, X and $\beta$ are defined as 
\begin{align*}
	X &=  \left( \begin{array}{ccccccccc}
1 & 1 & 0 & 1 & 0 & 1 & 0 & 0 & 0\\
1 & 1 & 0 & 0 & 1 & 0 & 1 & 0 & 0\\
1 & 0 & 1 & 1 & 0 & 0 & 0 & 1 & 0 \\ 
1 & 0 & 1 & 0 & 1 & 0 & 0 & 0 & 1 \end{array} \right) ~~~
	\beta =  \left( \begin{array}{c}
	\mu \\ \alpha_1\\ \alpha_2 \\ \beta_1 \\ \beta_2 \\ (\alpha \beta)_{11} \\ (\alpha \beta)_{12} \\ (\alpha \beta)_{21} \\ (\alpha \beta)_{22} \end{array} \right)
\end{align*}
The model can be expressed as
\begin{align*}
	\ln m_{ij} &= x_i^{'} \beta \\
	&= \mu + \alpha_i+ \beta_j + (\alpha \beta)_{ij}	
\end{align*}
We rename the parameters as 
\begin{align*}
	\ln m_{ij} &= u + u_{1(i)} + u_{2(j)} + u_{12(ij)}
\end{align*}
We notice that the above model is the same as the LLM defined for two-factor tables, where $u$ is the overall mean, $u_1, u_2$ are the main effects and $u_{12}$ is the interaction effect.

We note that we can fit LLMs as generalized linear models by using software packages available for generalized linear models, for example glm() function in ``stats" R package.
\subsection{Classes of Log Linear Models}
In this section, we discuss various classes of LLMs.
\subsubsection{Comprehensive Log-linear models}
The class of comprehensive Log-linear models is defined as follows. 
\begin{definition}[Comprehensive Log-linear Model] A log-linear model is said to be comprehensive if it contains the main effects of all the factors.
\end{definition}
For example, A Comprehensive LLM for the three-factor contingency tables must include all the main effects $u_{1}$, $u_2$ and $u_3$ along with other interaction effects if any, see \cite{Daniel} for the details.
\subsubsection{Hierarchical Log-linear models}
The class of hierarchical log-linear models is defined as follows.	 
\begin{definition}[Hierarchical Log-linear Models]
A LLM is said to be hierarchical if it contains all the lower-order terms which can be derived from the variables contained in a higher-order term. 
\end{definition}
For example, if a model for three-dimension table includes $u_{12}$ then $u_1$ and $u_2$ must be present or conversely if $u_2 = 0$ then we must have $u_{12} = u_{123} = 0$. 

It can be noted that hierarchical models may be represented by giving only the terms of highest order, also known as generating class, since all the lower-order terms are implicit. Generating class is defined as follows.
\begin{definition}[Generating class]
The highest order terms in hierarchical Log-linear models are called generating class because they generate all of the lower order terms in the model. 
\end{definition}
\begin{example}
A log linear model with generating class C = \{[123],[34]\} corresponds to the following log-linear model.
\begin{align*}
	 \ln m_{hijk} &= u + u_{1(h)} + u_{2(i)}+u_{3(j)} + u_{4(k)}+u_{12(hi)} + u_{23(ij)}+u_{13(hj)} +u_{123(hij)}+ u_{34(jk)}\\
	 &\text{members of generating class } [123] = \{[1],[2],[3],[12],[23],[13],[123]\}\\
	 &\text{members of generating class } [34] = \{[3],[4],[34]\}
\end{align*}
\end{example}
All models considered in the remaining sections of this article are hierarchical and comprehensive LLMs unless stated otherwise.

\subsubsection{Graphical Log-linear Models}
In this section, we consider a class of LLMs that can be represented by graphs, called graphical log-linear models(GLLMs).

\begin{definition}[Graphical Log-linear Models(GLLMs)]
A LLM model is said to be $graphical$ if it contains all the lower order terms which can be derived from variables contained in a higher-order term, the model also contains the higher order interaction.
\end{definition}
For example, if a model includes $u_{12}, u_{23}$ and $u_{31}$ it also contains the term $u_{123}$. \\
\\

In GLLMs, the vertices correspond to the factors and the edges correspond to the two-factor interactions. But the factors(vertices) and the two-factor interactions(edges) alone do not specify the graphical models. As mentioned previously, factorization of the probability distribution with respect to a graph must satisfy the Markov properties. 
For such a graph that respects Markov property with respect to a probability distribution, there is one-to-one correspondence between GLLMs and graphs. It follows that every GLLM determines a graph and every graph determines a GLLM, it is illustrated by the following examples.
\begin{example}
Let us consider the model [123][134]. The two factor terms generated by [123] are [12][13][23], similarly two factor terms generated by [134] are [13][14][34]. The corresponding graph is as given in the figure (\ref{figure:1}):

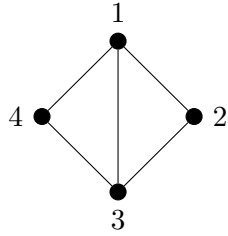
\begin{figure}[h!]
	\centering
\[\begin{tikzpicture}
	\vertex (A) at (1,2) [label=above:$1$] {};  	
	\vertex (B) at (2,1) [label=right:$2$] {};
	\vertex (C) at (1,0) [label=below:$3$] {};
	\vertex (D) at (0,1) [label=left:$4$] {};
	\path
		(A) edge (B)
		(A) edge (C)
		(A) edge (D)
		(B) edge (C)
		(C) edge (D)
	 ;   
\end{tikzpicture}\]
\caption{Graphical Model [123],[134]}
	\label{figure:1}
\end{figure}

Conversely we can also read log-linear model directly from the corresponding graph.
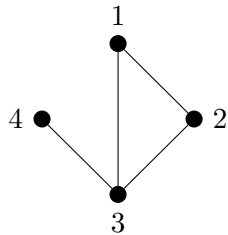
\begin{figure}[h!]
	\centering
\[\begin{tikzpicture}
	\vertex (A) at (1,2) [label=above:$1$] {};  	
	\vertex (B) at (2,1) [label=right:$2$] {};
	\vertex (C) at (1,0) [label=below:$3$] {};
	\vertex (D) at (0,1) [label=left:$4$] {};
	\path
		(A) edge (B)
		(A) edge (C)
		(B) edge (C)
		(C) edge (D)
	 ;   
\end{tikzpicture}\]
	\caption{Graphical Model [123] [34]}
	\label{figure:2}
\end{figure}

Consider a graph given in the figure(\ref{figure:2}), the edges are [12], [23], [13] and [34].  Since the generating class for the terms [12], [23] and [13] is the term [123], we must include [123] in the model. Hence the corresponding GLLM is [123][34].
\end{example}
We must note that, generating classes of graphical log-linear models are in a one-to-one correspondence with the maximal cliques of corresponding graph. Let us also note that not all hierarchical log-linear models have graphical representation. For example the model [12][13][23] is hierarchical but it is not graphical since it does not contain the higher order term [123].
\subsubsection{Decomposable Models}
In this section, we define the class of decomposable models which is a subclass of GLLMs. 
\begin{definition} [Decomposable log-linear Models]
A LLM model is decomposable if it is both graphical and chordal.
\end{definition}
The main advantage of this model over other models is that it has closed form Maximum Likelihood Estimates(abbreviated as MLE, which is explained in the later sections). 

For example, let us consider a decomposable model given by the figure (\ref{figure:1}). The only conditional independence implied by the graph is that, given the factors 1 and 3, factors 2 and 4 are independent. The maximum likelihood estimates for the expected cell counts are factorized in a closed form in the terms of sufficient statistics as
\begin{align*}
	\hat{m}_{hijk} = n_{hij.} n_{h.jk}/n_{h.j.}
\end{align*} 
The derivation of the above such expressions are discussed in the details in section 5.

The table (\ref{table:2}) shows all the possible non-isomorphic graphical and decomposable models for the four-factor contingency tables.

Few important articles concerned with the decomposable models are [\cite{Goodman2}, \cite{Goodman4}, \cite{Haberman1}, \cite{Haberman1}, \cite{Vijayan},\cite{Meeden98theadmissibility}, \cite{CorDecompModel}]. \\
\begin{small}
 \begin{longtable}{m{3cm} m{5cm} l}
	\caption{Graphical Log-linear Models for four-way tables} \label{table:2}\\
   \hline
    Model    &  $\quad \quad \quad $ Graph & Closed-form Estimates \\ \hline
    [1][2][3][4]   &  
    
	\begin{minipage}{.3\textwidth}
		\[\begin{tikzpicture}
			\vertex (A) at (1,2) [label=above:$1$] {};  	
			\vertex (B) at (2,1) [label=right:$2$] {};
			\vertex (C) at (1,0) [label=below:$3$] {};
			\vertex (D) at (0,1) [label=left:$4$] {};
			\path
	 		;   
		\end{tikzpicture}\]
     \end{minipage}
     & $\hat{m}_{hijk} = \frac{n_{h...} n_{.i..} n_{..j.} n_{...k}}{n_{....}^{3}}$\\
         
    [12][3][4]   &      
	\begin{minipage}{.3\textwidth}
		\[\begin{tikzpicture}
			\vertex (A) at (1,2) [label=above:$1$] {};  	
			\vertex (B) at (2,1) [label=right:$2$] {};
			\vertex (C) at (1,0) [label=below:$3$] {};
			\vertex (D) at (0,1) [label=left:$4$] {};
			\path
			(A) edge (B)
	 		;   
		\end{tikzpicture}\]
    \end{minipage}
    & 
    $\hat{m}_{hijk} = \frac{n_{hi..} n_{..j.} n_{...k}}{n_{....}^{2}}$ 		
    \\
    
    [12][13][4]   &      
	\begin{minipage}{.3\textwidth}
		\[\begin{tikzpicture}
			\vertex (A) at (1,2) [label=above:$1$] {};  	
			\vertex (B) at (2,1) [label=right:$2$] {};
			\vertex (C) at (1,0) [label=below:$3$] {};
			\vertex (D) at (0,1) [label=left:$4$] {};
			\path
			(A) edge (B)
			(A) edge (C)
	 		;   
		\end{tikzpicture}\]
    \end{minipage}
    &
     $\hat{m}_{hijk} = \frac{n_{hi..} n_{h.j.} n_{...k}}{n_{h...}n_{....}}$ 
     \\
    
    [12][34]   &      
	\begin{minipage}{.3\textwidth}
		\[\begin{tikzpicture}
			\vertex (A) at (1,2) [label=above:$1$] {};  	
			\vertex (B) at (2,1) [label=right:$2$] {};
			\vertex (C) at (1,0) [label=below:$3$] {};
			\vertex (D) at (0,1) [label=left:$4$] {};
			\path
			(A) edge (B)
			(C) edge (D)
	 		;   
		\end{tikzpicture}\]
    \end{minipage}
    & 
     $\hat{m}_{hijk} = \frac{n_{hi..} n_{..jk}}{n_{....}}$ 
     \\
    
    [12][13][14]   &      
	\begin{minipage}{.3\textwidth}
		\[\begin{tikzpicture}
			\vertex (A) at (1,2) [label=above:$1$] {};  	
			\vertex (B) at (2,1) [label=right:$2$] {};
			\vertex (C) at (1,0) [label=below:$3$] {};
			\vertex (D) at (0,1) [label=left:$4$] {};
			\path
			(A) edge (B)
			(A) edge (C)
			(A) edge (D)
	 		;   
		\end{tikzpicture}\]
    \end{minipage}
    &
     $\hat{m}_{hijk} = \frac{n_{hi..} n_{h.j.} n_{h..k}}{n_{h...}^{2}}$  
     \\    
    $[12][23][34]$ &      
	\begin{minipage}{.3\textwidth}
		\[\begin{tikzpicture}
			\vertex (A) at (1,2) [label=above:$1$] {};  	
			\vertex (B) at (2,1) [label=right:$2$] {};
			\vertex (C) at (1,0) [label=below:$3$] {};
			\vertex (D) at (0,1) [label=left:$4$] {};
			\path
			(A) edge (B)
			(B) edge (C)
			(C) edge (D)
	 		;   
		\end{tikzpicture}\]
    \end{minipage}
    & 
     $\hat{m}_{hijk} = \frac{n_{hi..} n_{.ij.} n_{..jk}}{n_{.i..}n_{..j.}}$ 
     \\
    
    [123][4]   &      
	\begin{minipage}{.3\textwidth}
		\[\begin{tikzpicture}
			\vertex (A) at (1,2) [label=above:$1$] {};  	
			\vertex (B) at (2,1) [label=right:$2$] {};
			\vertex (C) at (1,0) [label=below:$3$] {};
			\vertex (D) at (0,1) [label=left:$4$] {};
			\path
			(A) edge (B)
			(B) edge (C)
			(A) edge (C)
	 		;   
		\end{tikzpicture}\]	
    \end{minipage}
    & 
     $\hat{m}_{hijk} = \frac{n_{hij.} n_{...k}}{n_{....}}$ 
   \\
    
    [123][14]   &      
	\begin{minipage}{.3\textwidth}
		\[\begin{tikzpicture}
			\vertex (A) at (1,2) [label=above:$1$] {};  	
			\vertex (B) at (2,1) [label=right:$2$] {};
			\vertex (C) at (1,0) [label=below:$3$] {};
			\vertex (D) at (0,1) [label=left:$4$] {};
			\path
			(A) edge (B)
			(B) edge (C)
			(A) edge (C)
			(A) edge (D)
	 		;   
		\end{tikzpicture}\]
    \end{minipage}
    & 
    $\hat{m}_{hijk} = \frac{n_{hij.} n_{h..k}}{n_{h...}}$ 
    \\
    
    [12][23][34][14]   &      
	\begin{minipage}{.3\textwidth}
		\[\begin{tikzpicture}
			\vertex (A) at (1,2) [label=above:$1$] {};  	
			\vertex (B) at (2,1) [label=right:$2$] {};
			\vertex (C) at (1,0) [label=below:$3$] {};
			\vertex (D) at (0,1) [label=left:$4$] {};
			\path
			(A) edge (B)
			(B) edge (C)
			(C) edge (D)
			(A) edge (D)
	 		;   
		\end{tikzpicture}\]	
    \end{minipage}
    & No closed-form estimates exist \vspace{10pt} \\   
    
    [123][134]   &      
	\begin{minipage}{.3\textwidth}
		\[\begin{tikzpicture}
			\vertex (A) at (1,2) [label=above:$1$] {};  	
			\vertex (B) at (2,1) [label=right:$2$] {};
			\vertex (C) at (1,0) [label=below:$3$] {};
			\vertex (D) at (0,1) [label=left:$4$] {};
			\path
			(A) edge (B)
			(B) edge (C)
			(C) edge (D)
			(A) edge (C)
			(A) edge (D)
	 		;   
		\end{tikzpicture}\]		
    \end{minipage}
    &
     $\hat{m}_{hijk} = \frac{n_{hij.} n_{h.jk}}{n_{h.j.}}$  
    \\
    
    [1234]   &      
	\begin{minipage}{.3\textwidth}
		\[\begin{tikzpicture}
			\vertex (A) at (1,2) [label=above:$1$] {};  	
			\vertex (B) at (2,1) [label=right:$2$] {};
			\vertex (C) at (1,0) [label=below:$3$] {};
			\vertex (D) at (0,1) [label=left:$4$] {};
			\path
			(A) edge (B)
			(B) edge (C)
			(C) edge (D)
			(A) edge (C)
			(A) edge (D)
			(B) edge (D)
	 		;   
		\end{tikzpicture}\]		
    \end{minipage}
    & 
     $\hat{m}_{hijk} = n_{hijk}$  \vspace{10pt} \\    \hline	
\end{longtable}
\end{small}

\section{Statistical Properties of the Log-linear Models}
In this section, we discuss statistical properties of the hierarchical LLMs, like the existence of sufficient statistics, uniqueness of the MLE and model testing. First, we derive sufficient statistics for the unknown parameters of the model. Then we show how to compute the MLE of the expected cell counts from the sufficient statistics without computing the model parameters. We also show that for some models the estimated cell counts are the explicit closed function of the sufficient statistics, whereas for others we need iterative procedures.

\subsection{The Sufficient Statistics for LLMs}
We show that the sufficient statistics exist for the hierarchical LLMs and they are very easy to obtain. 
Let us consider the saturated model with simple multinomial sampling distribution for the 3-factor contingency tables. The log-likelihood function of the multinomial is obtained from the pdf given by the equation(\ref{eq:1}) as follows.
\begin{align} \label{eq:5}
	ln f(\{n_{ijk}\}) &=  \ln \left(\frac{N!}{ \prod_{ijk} n_{ijk}!} \right) + \sum_{ijk} {n_{ijk}} \ln (m_{ijk}) - N \ln N 	
\end{align}
Or equivalently we can write the above expression as
\begin{align} \label{eq:6}
	 ln f(\{n_{ijk}\}) &= \sum_{ijk} {n_{ijk}} \ln (m_{ijk}) 	+ C
\end{align}
Where $``C "$ represents the constant terms. Substituting the value for $\ln(m_{ijk})$ as given by the equation (\ref{eq:4}) we get the following expression.
\begin{align*}
	 ln f(\{n_{ijk}\}) &=  \sum_{ijk} {n_{ijk}} (u + u_1 +u_2+u_3+u_{12} + u_{13}+ u_{23}+ u_{123}) + C\\
	 &\text{the above expression can be also written as}\\
	 f(\{ n_{ijk}\}) &= exp( Nu +  \sum_i u_1 n_{i..} + \sum_i u_1 n_{i..} +  \sum_j u_2 n_{.j.} + \sum_k u_3 n_{..k} + \\
	 &\sum_{ij} u_{12} n_{ij.} + \sum_{ik} u_{13} n_{i.k} + \sum_{jk} u_{23} n_{.jk} + \sum_{ijk} u_{123} n_{ijk}  +C )
\end{align*}
Since multinomial distribution belongs to exponential family sufficient statistic exists, see \cite{AndersenEB}. From the above expression it is apparent that for the three-factor saturated model, the full table itself is the sufficient statistic since the lower order terms are redundant and it will be subsumed in the full table. 

We note that the marginal sub-tables which correspond to the set of generating classes are the sufficient statistics for the log-linear models, see \cite{Birch}.
\begin{example}
Consider a four-factors table with the following generating classes.
\begin{align*}
	\{ C_1, C_2  \} = \{ [123], [34]\}
\end{align*}
then $C_1(n)$ = $[n_{ijk.}]$ , it is a three-dimensional marginal sub table.\\
and $C_2(n)$ = $[n_{..kl}]$ , it is a two-dimensional marginal sub table.\\

These two marginal sub-tables are the sufficient for this model.
\end{example}
 For more details and the proofs on the sufficient statistics for hierarchical LLMs see \cite{Birch} and \cite{Haberman2}.
 
\subsection{The Maximum Likelihood Estimates for LLMs}
First, we state that a unique set of MLE for every cell count can be obtained from the sufficient statistics alone, see \cite{Birch} for the proof.
\\
\\
Now we state the Birch criteria as follows:\\
1. The marginal sub-tables obtained by summing over the factors not present in the max-cliques are the sufficient statistics for the corresponding expected cell counts. 
i.e., for the model \{[123], [34] \}, $C_1(n)$ = $((n_{ijk.}))$ and $C_2(n)$ = $((n_{..jk}))$ are sufficient statistics for $m_{ijk.}$ and $m_{..jk}$ respectively.\\
\\
2. All the sufficient statistics must be the same as the corresponding marginal sub-tables of their estimate means.
\begin{align*}
	C_i(\hat{m}) = C_i(n) ~~~~~~\forall i = 1  \text{ to \# of generating classes}
\end{align*}
i.e., for the model \{[123], [34] \} the estimated cell counts are
\begin{align*}
	\hat{m_{ijk.}} = n_{ijk.}\\
	\hat{m_{..kl}} = n_{..kl}
\end{align*}

Finally, the MLE of the expected cell counts for the model $\{[123], [34] \}$ is expressed as follows.
\begin{align*}
	\hat{m_{hijk}} = \frac{ n_{hij.} n_{..jk}} {n_{..j.}}\\
\end{align*} 
In section 5.4, we derive the closed form expressions for the MLEs in terms of sufficient statistics for three-factor contingency tables.

The reason for choosing MLE for computing the expected cell counts is its consistency and efficiency in the large samples. There is extensive research on the MLE of LLMs, we refer few of them here \cite{Glonek}, \cite{Andersen}, \cite{Haberman1}, \cite{Meeden98theadmissibility},\cite{Birch}, \cite{Stephen}, \cite{Lang2}, \cite{Lang6} and \cite{Darroch2} .
\subsection{Testing models}
The assessment of a model fit is very important as it describes how well it fits the data. We use the following test statistics.
\begin{itemize}
\item Pearson's $\chi^2$ Statistic:
which is defined as
\begin{align*}
	\chi^2 = \sum_i \frac{(O_i-E_i)^2}{E_i}
\end{align*}
where O denotes the observed cell counts and E as the expected cell counts.
\item Deviance goodness of fit test statistics:
We test a model against the saturated model using the deviance goodness of fit test, which is defined as follows.
\begin{align*}
	G^2 = -2 \sum_i  O_i \log \frac{E_i}{O_i}
\end{align*}
Under null hypotheses deviance is also distributes as $\chi^2$ with the appropriate degrees of freedom.
\end{itemize}

Table (\ref{table:3}), lists the degree of freedom of all the possible models for three-factor tables. For more information about the model testing we refer \cite{Davis}, \cite{Kreiner1} and \cite{Landis}.
\begin{table}[h]
\caption{Degrees of freedom} \label{table:3}
\centering
\begin{tabular}{c c}
    \toprule   
    Model   &  df \\     \toprule
    $[1][2][3] $  &   IJK - I - J - K + 2\\    
    $[12][3] $   &   (IJ-1)(K-1)\\    
    $[13][2] $   &   (IK-1)(J-1)\\    
    $[23][1] $   &   (JK-1)(I-1)\\    
    $[12][13]$    &   I(J-1)(K-1)\\    
    $[12][23]  $  &   J(I-1)(K-1)\\    
    $[13][23] $   &   K(I-1)(J-1)\\    
    $[12][13][23] $   &  (I-1)(J-1)(K-1)\\    
    $[123]$		& 0 \\    \bottomrule
\end{tabular}
\end{table}
\section{The Analysis of three-factor Contingency Tables}
In this section, we discuss the different interaction models for three-factor tables.
We also derive mathematical formulation for the MLE of the expected counts( when it is possible) for each model.
\subsection{The Complete Independence Model}
This is the simplest model where all the factors are mutually independent and $u_{12} = u_{13} = u_{23} = u_{123} = 0$. The following different equivalent notations can be used to represent this model.
\begin{align} \label{eq:7}
	 X \bigCI &Y \bigCI Z \nonumber\\	 
	 \ln(m_{ijk}) &= u+u1+u2+u3\\
	 C &= \{[1],[2],[3]\}\nonumber
\end{align}
This model can be represented graphically as given in the figure (\ref{figure:3}).
\begin{figure}[h!]
	\centering
\[\begin{tikzpicture}
	\vertex (A) at (1,2) [label=above:$1$] {};  	
	\vertex (B) at (2,1) [label=right:$2$] {};
	\vertex (C) at (1,0) [label=below:$3$] {};
	\path
		
	 ;   
\end{tikzpicture}\]
	\caption{The Complete Independence Model} \label{figure:3}
\end{figure}
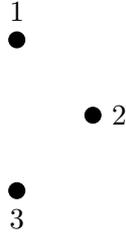
\begin{example}

When we substitute the value of $\ln (m_{ijk})$ as given in the equation (\ref{eq:4}) to log-likelihood kernel as given by the equation (\ref{eq:6}) and ignoring the constant term we get
\begin{align*}
	f(\{ n_{ijk}\}) &= \sum_{ijk} {n_{ijk}} \ln (m_{ijk}) 	\\
	&= \sum_{ijk} {n_{ijk}} (u + u_1 +u_2+u_3+)\\
	\text{ after simplification we obtain }\\
	 f(\{ n_{ijk}\}) &= exp( Nu +  \sum_i u_1 n_{i..} +  \sum_j u_2 n_{.j.} + \sum_k u_3 n_{..k} )
\end{align*}
From the above expression we obtain the sufficient statistics for this models as marginal sub-tables: $C_1 = \{n_{i..}\}$, $C_2 = \{n_{.j.}\}$ and $C_3 = \{n_{..j}\}$ which are estimates of $m_{i..}$, $m_{.j.}$ and $m_{..k}$ respectively.\\

From the equation (\ref{eq:7}), by summing over $jk, ik, ij$ and $ijk$ we obtain $m_{i..}$, $m_{.j.}$ and $m_{..k}$ and $m_{...}$ as 
\begin{align*}
 	{m_{i..}} &= exp(u+u_1) \sum_{jk} exp(u_2+u_3) \\
 	&= exp(u+u_1) \sum_{j} exp(u_2)\sum_k exp(u_3) \\
 	\\
	{m_{.j.}} &= exp(u+u_2) \sum_{ik} exp(u_1+u_3) \\
	&= exp(u+u_2) \sum_{i} exp(u_1)\sum_k exp(u_3)\\
	\\
	{m_{..k}} &= exp(u+u_3) \sum_{ij} exp(u_1+u_2) \\
	&= exp(u+u_3) \sum_{i} exp(u_1)\sum_j exp(u_2)\\
	\\
	{m_{...}} &= exp(u) \sum_{ijk} exp(u_1+u_2+u_3)\\
	 &= exp(u) \sum_{i} exp(u_1) \sum_{j} exp(u_2) \sum_{k} exp(u_3)
\end{align*}
	 
From the above equations we get the expression for $m_{ijk}$ as
\begin{align*}
		m_{ijk} &= \frac{ m_{i..} m_{.j.}m_{..k} }{m_{...}^2} 
\end{align*}
Applying Birch's result we get the estimates of $m_{ijk}$ as
\begin{align*}
	\hat{m_{ijk}} &= \frac{n_{i..} n_{.j.} n_{..k}}{n_{...}^2}
\end{align*}

Let us consider a contingency table as in table (\ref{table:1}), Under the complete independence assumption the sufficient statistics are the following marginal sub-tables.

\begin{center}
\makebox[0pt][c]{\parbox{1.2\textwidth}{
\begin{minipage}[b]{0.32\hsize}\centering
	
	\begin{longtable}[h!] {@{}cc@{}}  
	\caption{Personality Type}  \label{table:4}\\
	\centering		
		 Personality Type &\\
		\toprule 
		A &  1027 \\
		B & 1094 \\
		\bottomrule		
	\end{longtable}
\end{minipage}
\begin{minipage}[b]{0.32\hsize}\centering

	\begin{longtable}[h!] {@{}ccc@{}} 	
	\caption{Cholestrol}  \label{table:5}\\	
		\centering
		 Cholestrol & \phantom{ab}& \\
		\toprule 
		Normal & &1681 \\
		High && 440 \\
		\bottomrule
	\end{longtable}
\end{minipage}
\begin{minipage}[b]{0.32\hsize}\centering

	\begin{longtable}[h!] {@{}ccc@{}}
	\caption{DBP}  \label{table:6}\\ 		
		\centering
		 Cholestrol & \phantom{ab}& \\
		\toprule 
		Normal & &1005 \\
		High && 1116 \\
		\bottomrule
	\end{longtable}
\end{minipage}
}}
\end{center}

Under the complete independence assumption the table of fitted values are 
\begin{longtable}[h!] {@{}ccc|cc@{}} 
		\caption{Table of Estimated Cell Counts} \label{table:7}\\
		\toprule \centering
	 	&& &\multicolumn{2}{c}{Diastolic Blood Pressure}  \\
		\cmidrule{4-5}
		 Personality Type & \phantom{ab}& Cholestrol & Normal & High \\ \midrule
		
		A &&Normal& 739.9 &74.07\\
		& &High&    193.7 & 19.39 \\

		B &&Normal& 788.2 & 78.9\\
		& &High&    206.3 & 20.65\\
		\bottomrule
\end{longtable}
\end{example}
The $G^2$ statistic for the model is $8.723$(df:4, p-value:0.068), hence we conclude that the data supports the complete independence model. For details on Chi-Squared test of Independence we refer the reader to \cite{Goodman4}. 
\subsection{The Joint Independence Model}
\begin{figure}[h!]
	\centering
\[\begin{tikzpicture}
	\vertex (X) at (1,2) [label=above:$1$] {};  	
	\vertex (Y) at (2,1) [label=right:$2$] {};
	\vertex (Z) at (1,0) [label=below:$3$] {};
	\path
		(X) edge (Y);
	 ;   
\end{tikzpicture}\]
	\caption{The Joint Independence Model} \label{figure:4}
\end{figure}
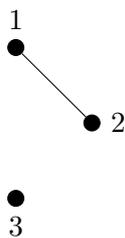

Under this model two factors are jointly independent of the third factor. There are three versions of this model, depending on which factor is unrelated to the other two. These three models are $(X ,Y) \bigCI Z$ , $(X ,Z) \bigCI Y$ and $(Y,Z) \bigCI X$. We consider only $(X ,Y) \bigCI Z$ in detail as others are comparable.\\
Equivalent different notations are as
\begin{align} \label{eq:8}
	 (X ,Y) &\bigCI Z \nonumber \\	 
	 \ln(m_{ijk}) &= u+u_{1}+u_{2}+u_{3}+u_{12}\\
	 C &= \{[12], [3]\} \nonumber
\end{align}
This model can also be represented graphically as given in the figure (\ref{figure:4}).
The sufficient statistics for this models are the marginal sub-tables: $C_1 = \{n_{ij.}\}$ and $C_2 = \{n_{..j}\}$  which are the estimates of $m_{ij.}$ and $m_{..k}$. From the equation (\ref{eq:8}) we obtain
\begin{align*}
 	{m_{ij.}} &= exp(u+u_1+u_2+u_{12}) \sum_{k} exp(u_3) \\
	{m_{..k}} &= exp(u+u_3) \sum_{ij} exp(u_1+u_2+u_{12}) \\
	 {m_{...}} &= exp(u) \sum_{ij} exp(u_1+u_2+u_{12}) \sum_{k} exp(u_3)
	 \\
\end{align*}
From the above equations we derive the closed form expression for $m_{ijk}$ as
\begin{align*}
	 m_{ijk} &= \frac{ m_{ij.} m_{..k} }{m_{...}}
\end{align*}
Applying Birch's criteria we get
\begin{align*}
	\hat{m_{ijk}} &= \frac{n_{ij.} n_{..k}}{n_{...}}
\end{align*}
We note that if the previous model of the complete independence ($X \bigCI Y \bigCI Z $) fits a data set, then the model ($(XY)\bigCI Z$) will also fit. But the smallest model will be preferred.\\

\begin{example} Let us consider another example to discuss this model.
\begin{longtable} {@{}ccc|cc@{}} 
		\caption{Classroom Behaviour Table (Everitt,1977)}  \label{table:8}\\
		\toprule \centering
	 	&& &\multicolumn{2}{c}{Risk}  \\
		\cmidrule{4-5}
		 Classroom Behaviour & \phantom{a}& Adversity of School  & Not at risk & at Risk \\ \midrule
		
		Nondeviant && Low & 16 & 7 \\
		& & Medium &    15 & 34 \\
		& & High &    5  & 3 \\
		
		Deviant && Low & 1 & 1 \\
		& & Medium &    3 & 8 \\
		& & High &    1 & 3 \\
		\bottomrule
\end{longtable}
\newpage
The sufficient statistics are
\begin{center}
\makebox[0pt][c]{\parbox{1.2\textwidth}{
\centering
\begin{minipage}[b]{0.32\hsize}\centering
	\begin{longtable}[h!] {@{}ccc@{}}  
	\caption{Adv*risk} \label{table:9}\\
	\toprule \centering	
		
		& \multicolumn{2}{c}{Risk}  \\
		\cmidrule{2-3}
		Adversity  & Not at risk & at Risk \\ 
		\toprule 
		Low &   17 & 8 \\
		Medium & 18 & 42 \\
		High & 6 & 6 \\
		\bottomrule		
	\end{longtable}
\end{minipage} 
~ ~ ~
\begin{minipage}[b]{0.32\hsize}\centering	
	\begin{longtable}[h] {@{}cc@{}}  
	\caption{Classroom Beha.} \label{table:10}\\
	\toprule \centering				
		 Classroom Beha. & Total \\
		\toprule 
		Nondeviant &  80 \\
		Deviant &  17 \\
		\bottomrule		
	\end{longtable}
\end{minipage}
}}
\end{center}
\textbf{\\ \\}
Under assumption of this model, the table of the expected cell counts is give in the table (\ref{table:11}).
\begin{longtable}[h] {@{}ccc|cc@{}} 
		\caption{Table of Estimated Cell Counts} \label{table:11}\\
		\toprule \centering
	 	&& &\multicolumn{2}{c}{Risk}  \\
		\cmidrule{4-5}
		 Classroom Behaviour & \phantom{a}& Adversity of School  & Not at risk & at Risk \\ \midrule
		
		Nondeviant && Low & 14.020 & 6.597	 \\
		& & Medium &    14.845 & 34.639 \\
		& & High &     4.948  &   4.948 \\
		
		Deviant && Low & 2.979 & 1.402 \\
		& & Medium &    3.154 & 7.360 \\
		& & High &    1.051 &  1.051 \\
		\bottomrule
\end{longtable}

The $G^2$ statistic for this model is $5.560$(df:5, p-value:0.351), hence we conclude that the data supports the joint independence model.

\end{example}
\subsection{The Conditional Independence Model}
Under this model two factor are conditionally independent given the third factor. There are three version for this model as well, these are $X  \bigCI Y | Z$, $X  \bigCI Z | Y$ and $Y  \bigCI Z | X$. We consider only $X \bigCI Y |Z$ in the detail as derivation for others are similar.\\

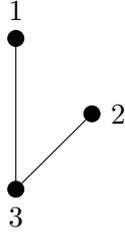
\begin{figure}[h!]
	\centering
\[\begin{tikzpicture}
	\vertex (X) at (1,2) [label=above:$1$] {};  	
	\vertex (Y) at (2,1) [label=right:$2$] {};
	\vertex (Z) at (1,0) [label=below:$3$] {};
	\path
		(X) edge (Z)
   		(Z) edge  (Y); 
	 ;   
\end{tikzpicture}\]
	\caption{The Conditional Independence Model} \label{figure:5}
\end{figure}

This model can be equivalently represented as
\begin{align}	\label{eq:9}
	 \ln(m_{ijk}) &= u+u_{1}+u_{1}+u_{3}+u_{13}+u_{23}\\
	 C&=\{[13][23]\} \nonumber
\end{align}
The graph for this model is given in the figure(\ref{figure:5}).
The sufficient statistics for this model are the marginal sub-tables: $C_{13} = \{n_{i.k}\}$ and $C_{23} = \{n_{.jk}\}$  which are estimates of $m_{i.k}$ and $m_{.jk}$. From the equation (\ref{eq:9}) we get
\begin{align*}
 	{m_{i.k}} &= exp(u+u_1+u_3+u_{13}) \sum_{j} exp(u_2+u_{23}) \\
	{m_{.jk}} &= exp(u+u_2+u_3+u_{23}) \sum_{i} exp(u_1+u_{13}) \\
	{m_{..k}} &= exp(u+u_3) \sum_{i} exp(u_1+u_{13}) \sum_{j} exp(u_2+u_{23})
\end{align*}
From the above three equations we obtain the closed form expression for $m_{ijk}$
\begin{align*}
	m_{ijk} &= \frac{ m_{ij.} m_{.jk} }{m_{..k}} 
\end{align*}
As before applying Birch's criteria we derive the expected counts for each cell as
\begin{align*}
	\hat{m_{ijk}} &= \frac{n_{ij.} n_{.jk}}{n_{..k}}
\end{align*}

\begin{example}
Let us consider the following infant's survival table, data taken from  \cite{Bishop1}.
\begin{longtable}[h!] {@{}ccc|cc@{}} 
		\caption{Infant Survival Table} \label{table:12}\\
		\toprule \centering
	 	&& &\multicolumn{2}{c}{Infant's Survival}  \\
		\cmidrule{4-5}
		 Clinic & \phantom{a}& Pre-natal care & Died & Survived \\ \midrule
		
		A && Less &    3  & 176\\
		& & More &     4 & 293\\

		B && Less &    17 & 197\\
		& & More &     2 & 23\\
		\bottomrule
\end{longtable}
\newpage

Assuming pre-natal care and survival is independent given clinic. The sufficient statistics are
\begin{center}
\makebox[0pt][c]{\parbox{1.2\textwidth}{
\centering
\begin{minipage}[b]{0.32\hsize}\centering
	
	\begin{longtable}[h] {@{}ccc@{}}  
	\caption{survival*clinic} \label{table:13} \\
	\toprule \centering	
		&\multicolumn{2}{c}{Infant's Survival}	\\
		\cmidrule{2-3}
		 Clinic & died & Survived\\
		\toprule 
		A &  7 & 469 \\
		B & 19  & 220\\
		\bottomrule		
	\end{longtable}
\end{minipage}
\begin{minipage}[b]{0.32\hsize}\centering	
	\begin{longtable}[h] {@{}ccc@{}}  
	\caption{clinic*prenatalcare} \label{table:14}\\
	\toprule \centering	
		&\multicolumn{2}{c}{Prenatal Care}	\\
		\cmidrule{2-3}
		 Clinic & Less & More\\
		\toprule 
		A &  179 & 297 \\
		B & 214  & 25 \\
		\bottomrule		
	\end{longtable}
\end{minipage}
\hfill
\begin{minipage}[b]{0.32\hsize}\centering
	
	\begin{longtable}[h] {@{}cc@{}} 
	\caption{clinic} \label{table:15}\\
	\toprule  \centering				
		 Clinic & Total \\
		\toprule 
		A & 476 \\
		B & 239\\
		\bottomrule		
	\end{longtable}
\end{minipage}
}}
\end{center}

\begin{longtable}[h] {@{}ccc|cc@{}} 
		\caption{Table of Estimated Cell Counts} \label{table:16}\\
		\toprule \centering
	 	&& &\multicolumn{2}{c}{Infant's Survival}  \\
		\cmidrule{4-5}
		 Clinic & \phantom{a}& Pre-natal care & Died & Survived \\ \midrule
		
		A && Less &     2.632  & 176.367 \\
		& & More &      4.367 &  292.632 \\

		B && Less &    17.012 & 196.987 \\
		& & More &     1.987 & 23.012 \\
		\bottomrule
\end{longtable}

The $G^2$ statistic for this model is $0.082$(df:2, p-value:0.959), hence we conclude that the data supports the conditional independence model.

\end{example}
\subsection{The Uniform Association Model}
This model is also known as no three-factor interaction model, where $u_{123} = 0$. For this model the log-linear notation is ([12], [13], [23]) but there is no graphical representation for this model. 
Unlike the previous models, there are no closed-form estimates for the expected cell counts/probabilities under this model. The maximum likelihood estimates can be computed by iterative procedures such as iterative proportional fitting(IPF) and Newton Raphson method.
\newpage
\begin{example}
Let us consider the following table taken from \cite{FienbergBook}.
\begin{longtable}[h] {@{}ccc|cc@{}} 
		\caption{Auto Accident Table} \label{table:17}\\
		\toprule \centering
	 	&& &\multicolumn{2}{c}{Injury}  \\
		\cmidrule{4-5}
		 Accident Type& \phantom{a}& Driver Ejected & Not Sever & Severe\\ \midrule
		
		Collision && No &    350  &  150 \\
		& & Yes &     26 & 23 \\

		RollOver && No &    60 & 112 \\
		& & Yes &     19 & 80 \\
		\bottomrule
\end{longtable}
None of the models discussed in previous sections fit the data. We use iterative proportional fitting algorithm to obtain the table of estimated counts as given in the table (\ref{table:18}).

\begin{longtable}[h] {@{}ccc|cc@{}} 
		\caption{Table of Estimated Cell Counts} \label{table:18}\\
		\toprule \centering
	 	&& &\multicolumn{2}{c}{Injury}  \\
		\cmidrule{4-5}
		 Accident Type& \phantom{a}& Driver Ejected & Not Sever & Severe\\ \midrule
		
		Collision && No &    350.48858 & 149.51130 \\
		& & Yes &     25.51142 & 23.48870 \\

		RollOver && No &    59.51104 & 112.48921 \\
		& & Yes &     19.48896 & 79.51079 \\
		\bottomrule
\end{longtable}

The $G^2$ statistic for this model is $0.043$(df:1, p-value:0.835), hence we conclude that the data supports the marginal association model.

For more information on IPF we refer to \cite{Deming} and \cite{StephenE}. We used the IPF procedure implemented in the R package ``cat", available at cran.r-project.org.
\end{example}

\subsection{The Saturated Model}
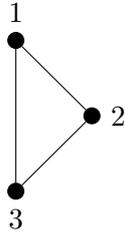
\begin{figure}[h!]
	\centering
\[\begin{tikzpicture}
	\vertex (X) at (1,2) [label=above:$1$] {};  	
	\vertex (Y) at (2,1) [label=right:$2$] {};
	\vertex (Z) at (1,0) [label=below:$3$] {};
	\path
		(X) edge (Z)
		(X) edge (Y)
   		(Z) edge  (Y); 
	 ;   
\end{tikzpicture}\]
	\caption{The Saturated Model} \label{figure:6}
\end{figure}

For this model the log-linear notation is ([123]). In this case there is no independence relationship between the three factors. The expected cell counts are the same as the observed cell frequencies.
\[
	 \hat{m}_{ijk}={n}_{ijk}
\]
Graphical representation for the saturated model is given in the fugure(\ref{figure:6}).

\begin{example}
Let us consider partial table which is based on clinical trial data \cite{Koch}.
\begin{longtable}[h!] {@{}ccc|ccc@{}} 
		\caption{Results of Clinical Trial for the Effectiveness of an
Analgesic Drug } \label{table:19}\\
		\toprule \centering
	 	&& &\multicolumn{3}{c}{Response}  \\
		\cmidrule{4-6}
		 Status & \phantom{a}& Treatment  &Poor & Moderate & Excellent \\ \midrule
		
		1 && Active & 3 & 20 & 5 \\
		& & placebo &    11 & 4 & 8 \\		
		
		2 && Active & 3 & 14 & 12 \\
		& & placebo &    6 & 13 & 5 \\
		\bottomrule
\end{longtable}
None of the model fits the data, we leave it for the reader to verify. 

\end{example}

\section{Model selection for the Decomposable Models}
In this section, we discuss model selection for the decomposable models only, since a non-decomposable graphical model  can be reduced to the decomposable one by adding minimal number of edges to the graph. For the details on minimum triangulation we refer to \cite{Tarjan} and an excellent survey article by \cite{Pinar}. 

Though decomposable models are a restricted family of GLLMs, selecting an optimal model from the class of decomposable graphical models is known to be an intractable problem. Most of all existing model selection algorithms are based on forward selection, backward elimination or combination of the both. There is a vast literature available for model selection and inference on graphical models, i.e., see \cite{Wainwright}, \cite{CorDecompModel} , \cite{Goodman3},  \cite{Ravikumar} and \cite{GenLLM}. 

We now illustrate the backward model selection procedure for a real data called ``women and mathematics(WAM)", data used in \cite{Fowlkes}. We use Wermuth's backward elimination algorithm, see \cite{Wermuth} for the details. The data is shown in the table(\ref{table:20}).
\newpage
\begin{longtable}[h!] {@{}llcccccccc@{}} 
		\caption{The table WAM }\label{table:20}\\
		\toprule \centering
		School& & \multicolumn{4}{c}{Suburban School}& \multicolumn{4}{c}{Urban School}  \\
		
		\toprule \centering
		Sex & &\multicolumn{2}{c}{Female}& \multicolumn{2}{c}{Male} &\multicolumn{2}{c}{Female}& \multicolumn{2}{c}{Male}  \\	 	
		\cmidrule{2-10}		
		
		Plan & Preference & Attend & Not & Attend & Not & Attend & Not & 	
		Attend  & Not \\
		\toprule \centering
		College & Maths-sciences \\
		&Agree & 37& 27& 51& 48& 51& 55& 109& 86\\
		& Disagree &  16& 11& 10& 19& 24& 28& 21& 25\\
		& Liberal arts\\
		&Agree & 16& 15&  7&  6& 32& 34& 30& 31\\
		& Disagree & 12& 24& 13&  7& 55& 39& 26& 19\\
		Job & Maths-sciences\\
		&Agree & 10&  8& 12& 15& 2& 1& 9& 5\\
		& Disagree & 9& 4& 8& 9& 8& 9& 4& 5\\
		&Liberal arts\\
		&Agree & 7& 10& 7& 3& 5& 2& 1& 3\\
		& Disagree & 8& 4& 6& 4& 10& 9& 3& 6 \\
		\bottomrule
\end{longtable}

Let us recall that graphical models are completely specified by their two-factor interactions. By the hierarchical principle, if a two-factor term is set to zero then any higher order term that contain that particular two-factor term will be also set to zero.

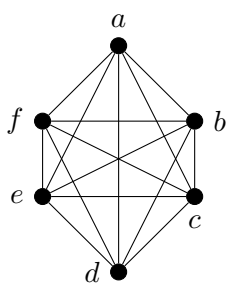
\begin{figure}[h!]
	\centering
	\[\begin{tikzpicture}
	\vertex (A) at (1,3) [label=above:$a$] {};  	
	\vertex (B) at (2,2) [label=right:$b$] {};
	\vertex (C) at (2,1) [label=below:$c$] {};
	\vertex (D) at (1,0) [label=left:$d$] {};
	\vertex (E) at (0,1) [label=left:$e$] {};
	\vertex (F) at (0,2) [label=left:$f$] {};
	\path
		(A) edge (B) (A) edge (C) (A) edge (D) (A) edge (E)
		(A) edge (F) 
		(B) edge (C) (B) edge (D) (B) edge (E) (B) edge (F)
		(C) edge (D) (C) edge (E) (C) edge (F)
		(D) edge (E) (D) edge (F) (E) edge (F)
	 ;   
	\end{tikzpicture}\]
\caption{The Saturated Model for WAM} \label{figure:7}
\end{figure}

The Wermuth's procedure starts with the saturated model, a single clique that includes all the two-factor effects as given in the figure (\ref{figure:7}). The vertices ``a",``b",``c",``d",``e",``f" correspond to the factors  ``Attendance",``Sex",``School", ``Agree", ``Subject" and ``Plans" respectively.

In the next step, all two-factor interactions $6 \choose 2$ are considered for elimination. We fix a backward elimination cut off level $\alpha = 0.5$. Among the two-factor interactions the term having the largest p-value are considered for elimination , only if the p-value exceeds $\alpha$. From the table (\ref{table:21}), we choose the edge (b,f) for deletion the resulting graphical model is the cliques [abcde][acdef] 
\begin{longtable}[h!] {@{}lllll@{}} 
		\caption{WAM: [abcde] } \label{table:21}\\
		\toprule \centering
		Edge & Clique & d.f. & $G^2$ & p-value  \\
		ab&[acdef][bcdef] & 16 & 18.585& 0.29078\\
		ac&[abdef][bcdef] & 16 & 20.689& 0.19080\\
		ad&[abcef][bcdef] & 16 &  14.172 & 0.58588\\
		ae&[abcdf][bcdef] & 16 & 18.781 & 0.28017\\
		af&[abcde][bcdef] & 16 & 11.951 & 0.74734 \\
		bc&[acdef][abdef] & 16 & 26.739 & 0.04447\\
		bd&[acdef][abcef] & 16 & 34.733 & 0.00432\\
		be&[acdef][abcdf] & 16 & 56.570 & 0.00000 \\
		bf&[acdef][abcde] & 16 & 11.673 & 0.76616\\
		cd&[abcef][abdef] & 16 & 29.439 & 0.02114\\
		ce&[abcdf][abdef] & 16 & 26.052 & 0.05329 \\
		cf&[abcde][abdef] & 16 & 81.657 & 0.00000\\
		de&[abcdf][abcef] & 16 & 78.248 & 0.00000\\
		df&[abcef][abcde] & 16 & 46.221 & 0.00009\\
		ef&[abcde][abcde] & 16 & 17.728 & 0.34005 \\
		\bottomrule
\end{longtable}

In the next step, we consider the cliques [abcde] and [acdef] . The edges ac, ad, ae, cd, ce and de are common to both the cliques, hence they are not considered for elimination. The candidate edges for deletion are ab, bc, bd, be, af, cf, df, ef. Let us examine the p-values for these edges as in the table(\ref{table:22}).

\begin{longtable}[h] {@{}lllll@{}} 
		\caption{WAM: [abcde][acdef]} \label{table:22}\\
		\toprule \centering
		Edge & Clique & d.f. & $G^2$ & p-value  \\
		ab& [bcde][acdef] & 8 &  12.456 & 0.13198 \\
		bc& [acde][acdef] & 8 & 18.097  & 0.02051\\
		bd& [acde][acdef] & 8 &  27.358  & 0.00061\\
		be& [acde][acdef] & 8 &   49.723 & 0.00000\\		
		af&[abcde][cdef] & 8 & 5.822 & 0.66711 \\		
		cf&[abcde][adef] & 8 & 73.014 & 0.00000\\		
		df&[abcde][acef] & 8 & 38.845 & 0.00001\\
		ef&[abcde][acdf] & 8 & 10.881 & 0.20852 \\
		\bottomrule
\end{longtable}

\begin{figure}[h!] 
\centering
	\[\begin{tikzpicture}
	\vertex (A) at (1,3) [label=above:$a$] {};  	
	\vertex (B) at (2,2) [label=right:$b$] {};
	\vertex (C) at (2,1) [label=below:$c$] {};
	\vertex (D) at (1,0) [label=left:$d$] {};
	\vertex (E) at (0,1) [label=left:$e$] {};
	\vertex (F) at (0,2) [label=left:$f$] {};
	\path
		(A) edge (B) (A) edge (C)  (A) edge (E)
		(B) edge (C) (B) edge (D) (B) edge (E) 
		(C) edge (D) (C) edge (E) (C) edge (F)
		(D) edge (E) (D) edge (F) (E) edge (F)
	 ;   
	\end{tikzpicture}\]
\caption{The Fitted Model for WAM} \label{figure:8}
\end{figure}
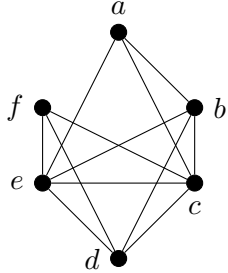

We delete the edges (af), the resulting graphical model is [abcde] 	[cdef]. Similarly we proceed further and in the next step the edge (ad) gets deleted and the resulting graphical model is [abce] [bcde] [cdef] as given in the figure(\ref{figure:8}).

In the next step candidate edges for deletion are [ab],[ac],[ae],[bd],[cf]. We notice that none of the p-value is greater than $\alpha = 0.05$ as given in the table (\ref{table:23}). So we stop with the model [abce][bcde][cdef].
 
\begin{longtable}[h!] {@{}lllll@{}} 
		\caption{WAM: [abce][bcde][cdef]} \label{table:23}\\
		\toprule \centering
		Edge & Clique & d.f. & $G^2$ & p-value  \\ \toprule
		ab& [ace] [bce][bcde][cdef] & 4 &  10.606 & 0.03137 \\
		ac& [bce] [ace][bcde][cdef] & 4 & 10.432 & 0.03374\\
		ae& [bce] [abc][bcde][cdef] & 4 & 10.426 & 0.03383\\
		bd& [abce] [cde][bce] [cdef] & 4 &  25.507 & 0.00004\\		
		cf&[abce] [bcde][def] [cde] & 4 &  67.832 & 0.00000\\				
		\bottomrule
\end{longtable}



\section{Computational details}
All the experimental results in this paper were carried out using R 3.1.3 . For fitting LLMs, there are several function in R, for example glm( ) and loglin( ) in the ``stats" library and loglm( ) in the "MASS" library. For model selection, we used dmod() and backward() functions implemented in the package ``gRim". All the packages used are 
available at http://CRAN.R-project.org/.

\section{Concluding Remarks}
In summary, we have discussed fundamental mathematical and statistical theory of GLLM and its applications. We restricted out attention to the complete table to make our discussion simple, as the tables having zero entries require special treatment. See chapter 8 of \cite{Christensen} for the analysis of contingency tables with zero cell counts. The limitations, and open problems in the use of GLLM for recursive relationships can be further explored, see section 5.4 of \cite{Christensen}. 
\newpage
\printbibliography

\end{document}